# On the rheology of pulmonary surfactant: effects of concentration and consequences for the surfactant replacement therapy

L.P.A. Thai, F. Mousseau, E.K. Oikonomou and J.-F. Berret[*]

*Matière et Systèmes Complexes, UMR 7057 CNRS Université Denis Diderot Paris-VII, Bâtiment Condorcet, 10 rue Alice Domon et Léonie Duquet, 75205 Paris, France.*

**Abstract:** The role of pulmonary surfactant is to reduce the surface tension in the lungs and to facilitate breathing. Surfactant replacement therapy (SRT) aims at bringing a substitute by instillation into the airways, a technique that has proven to be efficient and lifesaving for preterm infants. Adapting this therapy to adults requires to scale the administered dose to the patient body weight and to increase the lipid concentration, whilst maintaining its surface and flow properties similar. Here, we exploit a magnetic wire-based microrheology technique to measure the viscosity of the exogenous pulmonary surfactant Curosurf® in various experimental conditions. The Curosurf® viscosity is found to increase exponentially with lipid concentration following the Krieger-Dougherty law of colloids. The Krieger-Dougherty behavior also predicts a divergence of the viscosity at the liquid-to-gel transition. For Curosurf® the transition concentration is found close to the concentration at which it is formulated (117 g L$^{-1}$ *versus* 80 g L$^{-1}$). This outcome suggests that for SRT the surfactant rheological properties need to be monitored and kept within a certain range. The results found here could help in producing suspensions for respiratory distress syndrome adapted to adults. The present work also demonstrates the potential of the magnetic wire microrheology technique as an accurate tool to explore biological soft matter dynamics.

# I – Introduction

Complications related to premature births are the leading cause of death among children under 5 years of age, responsible for approximately 1 million deaths in 2015 [1-3]. Prematurity concerns infants born alive before the 37$^{th}$ week of pregnancy and represents nearly 1 in 10 births worldwide. During the last decades, significant progresses have been made in the management of neonates, in particular for the group of very preterm infants, those born before the 32$^{nd}$ week (representing 3.6% of all annual births). Changes in clinical practices have led to notable improvements in medical care, including controlled airway pressure, mechanical ventilation or *in situ* instillation of exogenous surfactant [3-5]. With regard to the latter, it is well known that very preterm neonates are lacking the pulmonary surfactant fluid lining the alveolar spaces. For these preterms, the absence of pulmonary surfactant results in strong breathing difficulties and eventually leads to a neonatal respiratory distress syndrome (RDS) [4,6-8]. The pulmonary surfactant is a complex surface-active fluid that contains phospholipids and proteins at a total concentration of about 35 - 40 g L$^{-1}$ [8-11]. Secreted by the Type II alveolar cells, this thin lipid film is essential for healthy lung functions as it reduces the surface tension between the air and the epithelium and also prevents alveolar collapse and expansion [12-15]. Surfactant replacement therapy (SRT) aims at bringing an exogenous surfactant by intratracheal instillation directly into the airways by forced inspiratory airflow. The most commonly used surfactants in neonatal and maternity hospitals are of porcine (Curosurf®) or bovine (Infasurf®, Survanta®, Alveofact®) origin. For the last decades, these drugs have been shown to be remarkably efficient in mitigating RDS.

Recent clinical studies have suggested that SRT could be also applied to children and adults suffering from similar syndromes [5,16,17]. SRT has been found to be beneficial in improving lung function in infants and children with respiratory failures, as for instance in case of meconium related airway obstruction and inflammation. For adults, similar strategies were attempted but recent clinical trials have shown mixed results [16,17]. These outcomes were in part explained by the fact that RDS pathologies are complex and involve multifactorial processes such as surfactant dysfunction, inflammation, cellular damages or edema [5].

On the biophysical side, major issues have been raised in adapting neonate oriented SRTs to adult lungs. The first issue is related to the upscaling of the applied dose from newborns to adults [5,6]. Typically, surfactant doses instilled intratracheally in premature infants are 200 mg/kg of body weight [18]. At the phospholipid concentrations of current clinical surfactants, this corresponds to a total volume of 2.5 mL/kg of body weight. In an adult of 80 kg, the instilled volume is increased to 200 mL, a value that is considered high for a medical intervention of this kind. This suggests the importance in modifying the existing exogenous formulations, either using biocompatible additives or concentrating the existing formulations to reduce the instilled volume [19,20]. The second issue has been put forward recently using fluid mechanics simulations dealing with the distribution of instilled surfactant in neonate and adult lungs [21,22]. It was shown that for a liquid bolus propagating through the airways by gravity, pulmonary surfactant displays a well-mixed distribution in the neonatal lungs, but poorly mixed and highly nonlinear distribution in adult lungs. The model suggests that adult surfactant therapy failures are likely due to inadequate delivery of the fluid in the lungs.

In this work, we focus on viscosity measurement of the exogenous pulmonary surfactant Curosurf® (*Chiesi Pharmaceuticals*, Parma, Italy). Curosurf® is a substitute widely used in neonatal and maternity hospitals for the treatment of RDS [4,12,23-26]. Our objective is first to provide accurate viscosity measurements of this lung mimicking fluid at clinically relevant concentrations. We also aim to explore whether the fluid displays viscoelasticity, an issue that has not been addressed yet and that could play an important role in the propagation of the liquid plug through the airways [21,22]. Pertaining to the pulmonary surfactant, bulk rheology measurements reported in the literature are scarce. King *et al*. have determined the shear rate dependent viscosity of endo- and exogenous (Infasurf®, Survanta®) surfactants between 100 and 800 $s^{-1}$ and have found shear-thinning behavior [27,28]. Lu and coworkers used a capillary viscometer to measure the viscosity in presence of polymers such as poly(ethylene glycol) or dextran. It was shown that the Infasurf® and Survanta® rheology can be finely tuned upon polymer addition [29]. In these studies, the viscosity determined at clinical concentrations was found in the range 5 – 50 mPa s, indicating a relatively low viscosity fluid [27-29]. The second objective is to determine the concentration dependence of the Curosurf® viscosity and to assess whether the existing formulations can be further concentrated for SRT and still remain of low viscosity.

Here, we exploit a recent microrheology technique dubbed magnetic rotational spectroscopy (MRS) based on the tracking of magnetic wires dispersed in a medium [30-33]. Submitted to an external magnetic field at increasing frequency, the viscosity is derived from the identification of the different wire rotation regimes. The MRS technique is relevant as it requires only a few microliters of sample and. Curosurf® dispersion containing lipid vesicles with a broad size distribution, the MRS technique also takes advantage of the fact that the wires have lengths ranging from 10 to 100 µm, therefore larger than the fluid constituents. With these measurements, we demonstrate that Curosurf® suspensions display an exponential increase of the viscosity with concentration, and discuss its consequences for surfactant replacement therapy.

# II – Materials and Methods

**II.1 – Materials** Curosurf® (*Chiesi Pharmaceuticals*, Parma, Italy) is a porcine minced pulmonary surfactant extract used in neonatal and maternity hospitals mainly in Europe [4]. It is produced as a 80 g L$^{-1}$ suspension containing among others phosphatidylcholine (PC) lipids, sphingomyelin (SM), phosphatidylethanolamine (PE)), phosphatidylinositol (PI), phosphatidylglycerol (PG) and the hydrophobic proteins SP-B and SP-C [23,34,35]. Its composition is compared to that of native surfactant in **Supplementary Information S1.** Curosurf® is administered by intratracheal instillation to premature newborns with respiratory distress syndrome. The optimum dose of 200 mg per kilogram body weight is given in a single bolus by highly trained and experienced staffs [18]. Three milliliters of Curosurf® 80 g L$^{-1}$ are worth about 1000 €. Exogenous surfactant such as Curosurf® is considered as a model fluid because its composition and physical properties are subject to little changes from one formulation to another and because of its long-term stability. Curosurf® was kindly provided by Dr. Mostafa Mokhtari and his team from the neonatal service at Hospital Kremlin-Bicêtre, Val-de-Marne, France. The suspension appears as a whitish and low viscosity fluid. Its gel-to-fluid transition temperature was estimated experimentally at 29.5 °C (**S2**). Curosurf® was used as received or diluted in phosphate buffer saline (PBS, Aldrich).

Iron oxide nanoparticles (γ-Fe$_2$O$_3$) were obtained by co-precipitation of iron(II) and iron(III) salts in aqueous solution according to the Massart synthesis [36]. The particle size (6.8 nm) and dispersity (0.18) were measured by TEM, whereas the maghemite cubic structure was assessed by electron beam diffraction [37] (**S3**). γ-Fe$_2$O$_3$ was then coated with poly(acrylic) acid polymers (PAA$_{2K}$, Aldrich, $M_w$ = 2100 g mol$^{-1}$) following the precipitation-redispersion method [38]. Wires were synthesized by electrostatic co-assembly with poly(diallyldimethylammonium chloride) (PDADMAC, Aldrich, $M_w$ > 100000 g mol$^{-1}$) [39] (**S3**). The magnetic wires used in this study have lengths between 10 and 100 μm and diameters between 1 and 3 μm. With the 100× objective (numerical aperture 1.3, lateral resolution 220 nm), we have found that the diameter correlates with the length, according to $D(L) = 0.619 L^{0.202}$ (**S4**). The previous equation is used for data analysis, in particular for microscopy experiments performed at 20×.

**II.2 – Dynamic Light scattering** The hydrodynamic diameter $D_H$ of Curosurf® vesicles was determined on dilute solutions using a NanoZS Zetasizer (Malvern Instruments). The second-order autocorrelation function was recorded in triplicate (T = 25 °C) and analyzed using the cumulant and CONTIN algorithms to determine the average diffusion coefficient (**S5**).

**II.3 – Electrophoretic mobility and zeta potential** Laser Doppler velocimetry using the phase analysis light scattering mode was performed using the Zetasizer Nano ZS equipment (Malvern Instruments, UK) to determine the electrophoretic mobility and zeta potential. At physiological pH, the Curosurf® vesicles and magnetic wires were found to be negatively charged, with zeta potential values $\zeta$ = -54 ± 8 mV [24] and $\zeta$ = -26 ± 4 mV [39], respectively.

**II.4 – Nanoparticle Tracking Analysis (NTA)** NTA measurements were performed with a NanoSight LM14 (Malvern Instruments, UK), equipped with a sample chamber illuminated by a 532-nm laser. Liquid samples were injected in the measuring chamber until the liquid reached the tip of the nozzle. The software used for recording and analyzing the data was Nanosight NTA 3.0. NTA data from Curosurf® vesicles are shown in **S6.**

**II.5 – Cryo-transmission electron microscopy (cryo-TEM)** For cryo-TEM, few microliters of a 5 g L$^{-1}$ Curosurf® dispersion were deposited on a lacey carbon coated 200 mesh (Ted Pella Inc.). The drop was blotted with a filter paper using a FEI Vitrobot$^{TM}$ freeze plunger. The grid was then quenched rapidly in liquid ethane to avoid crystallization and later cooled with liquid nitrogen [40]. The membrane was transferred into the vacuum column of a JEOL 1400 TEM microscope (120 kV).

**II.6 – Phase-contrast and bright field optical microscopy** Phase-contrast and bright field images were acquired on an IX73 inverted microscope (Olympus) equipped with 20× and 100× objectives. An EXi Blue camera (QImaging) and Metaview software (Universal Imaging Inc.) were used as acquisition system.

**II.7 – Rheology** Rheology experiments were performed using a Physica RheoCompass MCR 302 (Anton Paar) working with a cone-and-plate geometry (diameter 50 mm, cone angle 1°, sample volume 0.7 mL). The MCR 302 rheometer was operated in rate-controlled mode for the measurements of the storage and loss moduli $G'(\omega)$ and $G''(\omega)$ and of the stress *versus* shear rate curves $\sigma(\dot{\gamma})$. The macrorheology were carried out in duplicate and at two temperatures, T = 25 °C and 37 °C (**S7**).

**II.8 – Active microrheology** The magnetic wire microrheology technique has been described in previous accounts [41-43]. For reviews on microrheology techniques and data analysis, especially those using anisotropic probes, we refer to Refs [30,31,33,44]. Curosurf® stock suspensions were used as received and diluted to the desired concentrations of 5, 20, 40, 50, 70 and 80 g L$^{-1}$ using PBS. A total of 10$^5$ wires (contained in 0.5 µL) was then added to 100 µL of the previous suspensions and gently stirred. 25 µL of the previous suspension were deposited on a glass plate and sealed into to a Gene Frame® (Abgene/Advanced Biotech, dimensions 10×10×0.25 mm$^3$). The microrheology protocol is based on the Magnetic Rotational Spectroscopy (MRS) technique. MRS consists in applying a rotating magnetic field to a wire and recording its motion by time-lapse microscopy [30-32] (**S8**). For calibration, MRS was performed on a series of water-glycerol mixtures of increasing viscosities, 4.95, 34.9, 48.9 and 80.0 mPa s, corresponding to glycerol concentrations of 49.8%, 81.0%, 84.5% and 89% (T = 25 °C) (**S9**).

# III – Results and discussion
### III.1 – Curosurf® structure
Curosurf® nano- and micro-structure was resolved using a combination of techniques including cryo-TEM, dynamic light scattering, nanoparticle tracking analysis and phase contrast optical microscopy. Figs. 1a and **S10** display representative cryo-TEM images of Curosurf® diluted at 5 g L$^{-1}$. Images show that the phospholipids associate locally into bilayers with a thickness of 4.4 nm (**S11**). On a larger scale, the bilayers close on themselves and form unilamellar and multivesicular vesicles. Multivesicular vesicles describe large membrane compartments encapsulating one or several smaller vesicles. Spherical micelles made from these lipids were not observed. The vesicle size distribution obtained from cryo-TEM (Fig. 1b) is well described by a log-normal function of median diameter 230 nm and dispersity 0.55 (continuous line). The analysis was made considering all discernable vesicles ($n$ = 187), *i.e.* including those internalized in larger ones. More generally, cryo-TEM Curosurf® vesicles are similar to those reported in the literature [23,26,45] and comparable to those of endogenous surfactants [46]. Nanoparticle tracking analysis (NTA) performed on a $10^{-3}$ g L$^{-1}$ dispersion revealed a vesicle number distribution peaked around 150 nm, in fair agreement with the cryo-TEM results (**S6**). From the NTA measurements, the vesicle number density was estimated at $7.4 \times 10^{11}$ L$^{-1}$ and the volume fraction $\phi$ derived. We found the relationship $\phi = 5.2 \times 10^{-3} c$, where $c$ is given in g L$^{-1}$. In practice this means that the concentration at which Curosurf® is formulated, 80 g L$^{-1}$ is associated with the volume fraction of $\phi = 0.42$. Dynamic light scattering performed in the dilute regime at 1 g L$^{-1}$ shows a bimodal distribution with two major peaks at 160 nm and 700 nm (**S5**) [24,25,47]. Because light scattering is sensitive to large objects, the technique emphasizes the large vesicle population which is not directly identified from cryo-TEM and NTA experiments. Figs. 1c and 1d display a 20× magnified image of a 40 g L$^{-1}$ dispersion obtained from phase-contrast optical microscopy and the associated vesicle size distribution ($n$ = 332), respectively. With microscopy, vesicles are found in the range 1 – 10 µm with a median value at 3.3 µm. As for cryo-TEM, the vesicle size distribution is well adjusted by a log-normal function (continuous line in Fig. 1d). In conclusion, Curosurf® dispersions contain mostly lipids that assemble into vesicles. Their size distribution is broad and extends from 100 nm to 10 µm. The different techniques used have underscored the existence of different populations, which are composed of 100 – 200 nm mainly uni-lamellar vesicles, as observed by cryo-TEM and NTA and micron-sized multi-vesicular vesicles as shown in DLS and optical microscopy.

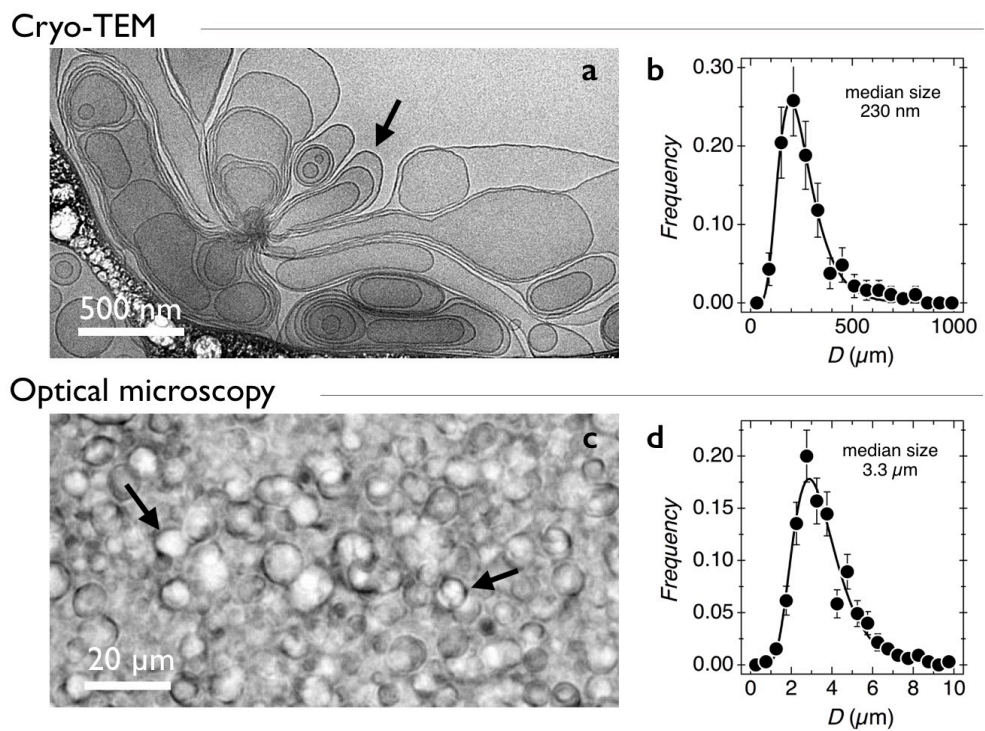

*Figure 1: a) Cryo-TEM image of native Curosurf® obtained from a 5 g L$^{-1}$ dispersion at 25 °C. The arrow is pointing to the vesicular membrane. b) Vesicle size distribution determined from cryo-TEM images. c) Phase-contrast optical microscopy image of a 40 g L$^{-1}$ Curosurf® dispersion at 25 °C (objective 20×). Arrows are pointing to vesicles. d) Vesicle size distribution determined from optical microscopy images (median size 3.3 µm, dispersity 0.4).*

### III.2 – Curosurf® cone-and-plate rheology

Cone-and-plate rheology was monitored on Curosurf® suspensions to serve as a reference for the microrheology experiments. To reduce the amount of sample used, measurements were performed at two concentrations, 40 and 80 g L$^{-1}$. Fig. 2a and 2b compare the rheological behaviors of Curosurf® at such concentrations in the linear and non-linear flow regimes. The experiments were performed in controlled shear rate mode at 25 °C and 37 °C (**S7**). In dynamical frequency sweeps, the loss moduli $G''(\omega)$ measured on two different samples are found to agree well with each other and to increase with the frequency according to a scaling of the form $G''(\omega) \sim \omega^\alpha$, where the exponent $\alpha$ equals 0.95 ± 0.03 at 40 g L$^{-1}$ and 0.75 ± 0.03 at 80 g L$^{-1}$. The storage modulus $G'(\omega)$ was also measured as a function of the frequency. However, the data were not considered reliable, the corresponding elastic torque being below the detection limit. The linear rheology results suggest that the 40 g L$^{-1}$ Curosurf® dispersion is a purely viscous Newton fluid (for which the exponent $\alpha = 1$), whereas the 80 g L$^{-1}$ exhibits some viscoelastic behavior.

In steady shear (Fig. 2b), the stress *versus* shear rate curve was measured following a decreasing ramp of velocity gradients starting from $\dot{\gamma}$ = 1000 down to 1 s$^{-1}$ [48]. Shearing below 1 s$^{-1}$ yields spurious behaviors which we again attribute to the lack of sensitivity of the rheometer. The non-linear stress data confirm the purely viscous character of the 40 g L$^{-1}$ sample, whereas the 80 g L$^{-1}$ exhibits a slight shear-thinning behavior [28]. The static viscosity $\eta$ was extrapolated from the flow curve at the shear rate of 0.01 s$^{-1}$ and found to be 4.0 ± 0.5 mPa s and 20.0 ± 1.0 mPa s for the 40 and 80 g L$^{-1}$ samples, respectively (**S7**). The viscosity values obtained for Curosurf® at physiological conditions are in the same range (~ 5 mPa s) as those determined by a cone-and-plate rheometry on endogenous surfactant [28] and by capillary viscometry on Infasurf® at 35 g L$^{-1}$ [29]. They are, however lower than those obtained on Infasurf® and Survanta® in the same conditions [27-29]. In these later measurements, the viscosity reaches the values up to 50 mPa s. In summary, we have found that under physiological conditions ($c$ = 40 g L$^{-1}$) Curosurf® is a low viscosity fluid, with a viscosity coefficient slightly above that of the solvent. As the concentration is increased to that of the formulation ($c$ = 80 g L$^{-1}$), the viscosity exhibits a five-time increment and the fluid displays weak viscoelasticity and shear-thinning behavior.

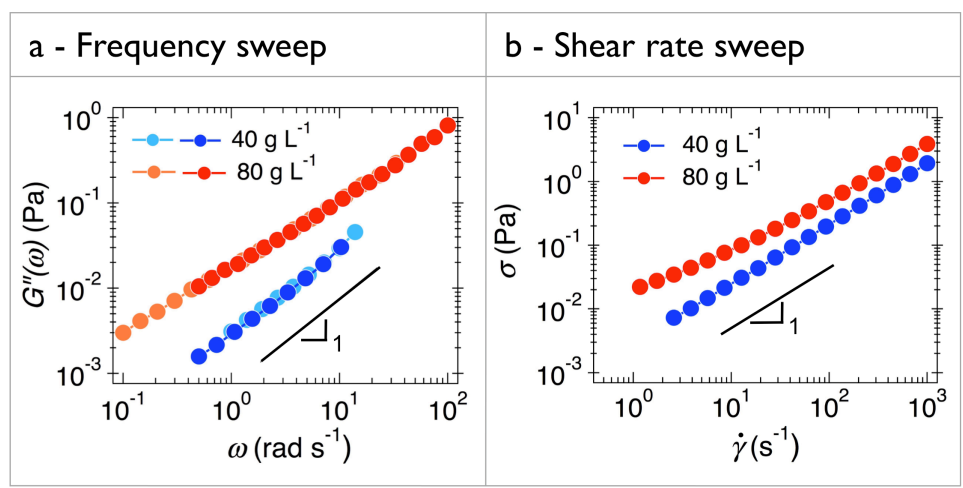

*Figure 2: a) Loss modulus $G''(\omega)$ measured for 40 and 80 g L$^{-1}$ Curosurf® suspensions using a rotational strain-controlled rheometer operating with a cone-and-plate (T = 25 °C). b) Shear stress versus shear rate curves for the samples of Fig. 2a (**S7**).*

### III.2 – Curosurf® magnetic wire microrheology
*III.2.1 – Transient behavior observed by MRS*

Here we provide an illustration of the time dependent wire motion in Curosurf® under the application of a rotating field. To this aim, we focus on a 94 µm wire embedded in a 40 g L$^{-1}$ suspension (T = 25 °C) and exposed to a driving field of amplitude $\mu_0 H$ = 10.3 mT. For each frequency tested, a 60 s image stack is recorded and digitalized. The position of the center-of-mass and orientation angle of the wire are then retrieved as a function of the elapsed time. Fig. 3a shows the wire orientation at different times (t = 14.7 s, 16.6 s, 18.5 s and 20.4 s) after the field inception.

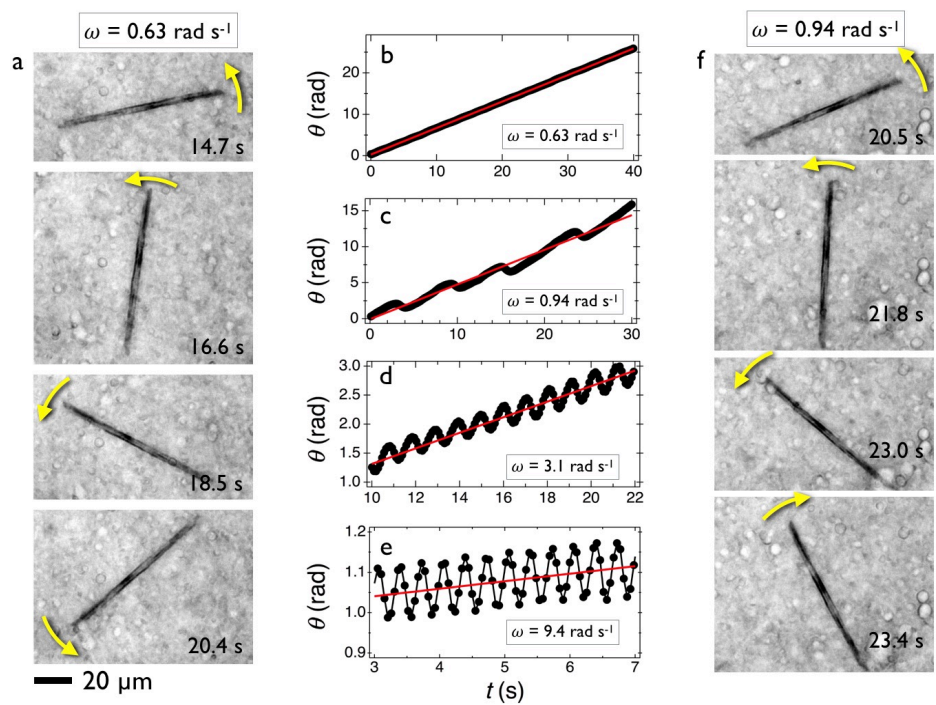

*Figure 3: a)* *Optical microscopy images of a 94 µm wire in Curosurf® 40 g L$^{-1}$ subjected to a rotating field of 10.3 mT at the frequency of 0.63 rad s$^{-1}$ (T = 25 °C). The arrows indicate a steady counter-clockwise rotation.* *b)* *Time dependence of the angle $\theta(t)$ corresponding to the experiment in a). The straight line in red is calculated from $\theta(t) = \omega t$, indicating synchronous rotation (**Movie#1**).* *c, d and e)* *Same representation as in b) for angular frequencies 0.94, 3.1 and 9.1 rad s$^{-1}$, respectively. These frequencies are in the non-synchronous regime ($\omega_C$ = 0.75 rad s$^{-1}$) where the wire displays back-and-forth oscillations.* *f)* *The microscopy images on the right-hand side show that after a counter-clockwise rotation, the wire comes back rapidly in a clockwise motion, indicating that the wire rotation is hindered (**Movie#2**).*

At the angular frequency of 0.63 rad s$^{-1}$, the wire rotates counter-clockwise in a regular manner, as indicated by the arrows. The movie provided in **SI** shows that Curosurf® vesicles are being stirred by means of the wire rotation and their displacement attests of the effective shearing of the vesicle continuum. Fig. 3b illustrates the corresponding time dependence of the angle $\theta(t)$ at this frequency. There, $\theta(t)$ increases linearly with time according to $\theta(t) = \omega t$, indicating that the wire rotates synchronously with the field. As the angular frequency is increased, the wire enters a second regime characterized by back-and-forth oscillations. The $\theta(t)$-traces displayed in Figs. 3c-3e were obtained at $\omega$ = 0.94, 3.1 and 9.4 rad s$^{-1}$. Their main features are a doubling of the oscillation frequency and a diminution of the oscillation amplitude. The transients are characterized by an average rotation frequency $\Omega(\omega) = \langle d\theta(t)/dt \rangle_t$ shown in the figures as straight lines in red. The transition between synchronous (I) and asynchronous (II) regimes can be illustrated using a series of snapshots taken at the actuating frequency, $\omega$ = 0.94 rad s$^{-1}$ (Fig. 3f). The images show that after a counter-clockwise rotation in the first 3 panels (between 20.5 and 23.0 s after the field inception), the wire comes back rapidly by 40 degrees in a clockwise motion, indicating that the wire does not follow the field in this time interval. The angular frequency at which the transition occurs reads [41,43] :

$$\omega_C = \frac{3}{8}\frac{\mu_0 \Delta\chi}{\eta} g\left(\frac{L}{D}\right)\frac{D^2}{L^2} H^2 \qquad (1)$$

where $\mu_0$ is the permeability in vacuum, $L$ and $D$ the length and diameter of the wire, $H$ the magnetic excitation amplitude and $g(L/D) = ln(L/D) - 0.662 + 0.917 D/L - 0.050 (D/L)^2$ is a dimensionless function of the anisotropy ratio [41,42]. In Eq. 1, $\Delta\chi = \chi^2/(2+\chi)$ denotes the anisotropy of susceptibility between parallel and perpendicular directions and $\chi$ represents the material magnetic susceptibility ($\Delta\chi = 0.056 \pm 0.006$, **S9**). For the wire in Fig. 4, the critical frequency $\omega_C$ was estimated at 0.75 rad s$^{-1}$, corresponding to a static viscosity $\eta$ of $6.9 \pm 1.7$ mPa s.

*III.2.2 – Angular frequency dependence*
In this section, the objective is to compare Curosurf® microrheology results with constitutive model predictions and at the same time to assess the validity of the MRS technique for vesicle suspensions. To this aim, the transient tracking experiment shown previously was replicated on five additional wires of length 27, 43, 64, 67 and 72 μm. The wire rotational behavior was monitored *versus* $\omega$ between $10^{-2}$ and $10^2$ rad s$^{-1}$ and the $\theta(t)$-traces were analyzed and translated into the average rotation velocity $\Omega(\omega, L)$. In parallel, the critical frequency $\omega_C(L)$ was determined for each wire. Both $\Omega(\omega, L)$ and $\omega_C(L)$ are important quantities because their theoretical expressions are known and their asymptotic behaviors have been derived for basic rheological fluids and solids [41,43,49]. Fig. 4a displays the average rotation velocity for the six wires expressed in reduced units, $\tilde{\Omega}(\omega) = \Omega(\omega)/\omega_C$ against $\tilde{\omega} = \omega/\omega_C$. With increasing frequency, the average velocity $\tilde{\Omega}(\omega)$ increases linearly, passes through a maximum at $\tilde{\omega} = 1$ and then decreases as $\tilde{\omega}^{-1}$. The data in Fig. 4a were adjusted using constitutive equations derived for viscous and viscoelastic liquids [41,42,49] :

$$\text{Regime I:} \quad \tilde{\omega} \leq 1 \quad \tilde{\Omega}(\tilde{\omega}) = \tilde{\omega}$$

$$\text{Regime II:} \quad \tilde{\omega} \geq 1 \quad \tilde{\Omega}(\tilde{\omega}) = \tilde{\omega} - \sqrt{\tilde{\omega}^2 - 1} \qquad (2)$$

In the synchronous regime, the agreement between the data and the model is excellent, whereas in the asynchronous regime the data are scattered apart from the model prediction (Eq. 2, continuous line in red). The good agreement observed in Regime I is due to the fact that the rotation speed does not depend directly on the fluid viscosity, in contrast to the situation in Regime II [49]. The continuous lines in grey were obtained by varying the critical frequency by ± 40% with respect to $\omega_C$. The discrepancy observed in the high frequency branch can be better visualized using the average rotation frequency residuals, $\Delta\tilde{\Omega}/\tilde{\Omega}$. There, $\Delta\tilde{\Omega}/\tilde{\Omega}$ is obtained by subtracting the prediction values to the experimental ones and by dividing with the prediction values. The deviations from the model observed in Fig. 4b are assigned to spatial heterogeneities of the vesicular fluid. These inhomogeneities are associated with concentration or interaction fluctuations at the scale of the probe. In some sequences, it can be seen that direct wire-vesicle interactions deflect the wire center of mass from a few microns or modify slightly its oscillation frequency (**Movie#3**). Such perturbations lead to fluctuations in the average rotation frequency $\tilde{\Omega}(\tilde{\omega})$ and to uncertainties in the determination of the viscosity. To support this conclusion, we have appended in **S12** results obtained for a viscoelastic fluid that is homogeneous at the scale of the wires. In such a case, $\tilde{\Omega}(\tilde{\omega})$ follows precisely the model predictions of Eq. 2 in the two regimes. For the pulmonary surfactant, as most of the data points are comprised between the ± 40% calculated curves, it can be concluded that $\omega_C$ and hence the viscosity are determined with an accuracy of about ± 40%.

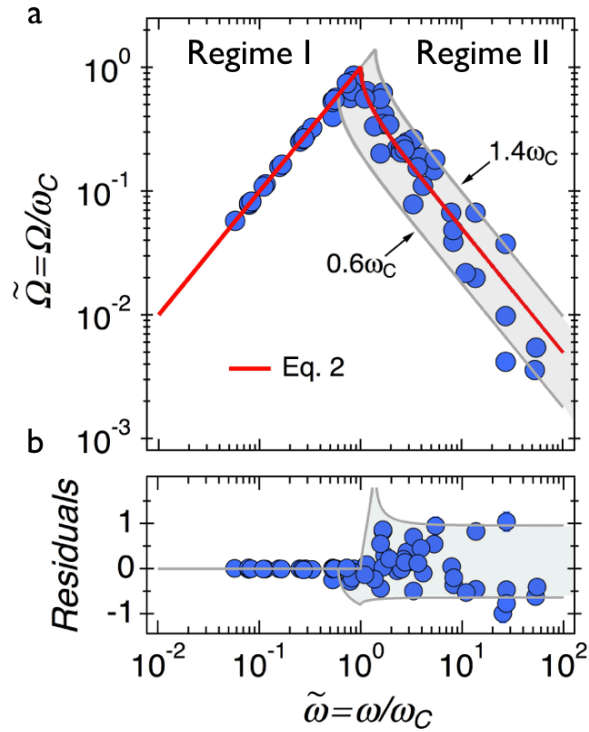

**Figure 4: a)** Wire average rotation velocity measured on a 40 g L$^{-1}$ Curosurf® suspension and expressed in reduced units, $\tilde{\Omega}(\omega) = \Omega(\omega)/\omega_C$ versus $\tilde{\omega} = \omega/\omega_C$ (T = 25 °C). The continuous line in red (from Eq. 2) indicates the transition between the synchronous and asynchronous regimes. The thin lines in grey were obtained by varying the critical frequency by ± 40%. **b)** $\Delta\tilde{\Omega}/\tilde{\Omega}$ residuals calculated from the data in **a)**.

*III.2.3 – Concentration dependences*
We turn now to the concentration dependence of the Curosurf® viscosity determined by MRS. To circumvent the accuracy issue mentioned previously, the critical frequency $\omega_C$ was determined experimentally as a function of the wire length for Curosurf® 0, 5, 20, 40, 50, 70 and 80 g L$^{-1}$. Data on the average rotational frequency $\Omega(\omega, L)$ and on the amplitude oscillation $\theta_B(\omega, L)$ obtained for each wire were treated consistently in close relation with constitutive model predictions [41]. This approach resulted in the data of Fig. 5a, where $\omega_C$ is shown as a function of the reduced length $L^*(L, D) = L/D\sqrt{g(L/D)}$ at increasing concentrations. In all studied examples, the $1/L^{*2}$-behavior was obtained, in agreement with Eq. 1. Compared to the previous measuring protocol, the values of the static viscosity show now a 20% relative error, instead of the previous 40%. Table I emphasizes moreover the excellent agreement between the macro- and microrheology viscosities obtained for the 40 and 80 g L$^{-1}$ suspensions. In Fig. 5b, the static viscosity $\eta(c)$ is found to increase linearly at low concentration (discontinuous straight line) and to deviate significantly from the linear behavior above ~ 40 g L$^{-1}$. The continuous line through the data points was obtained using the Krieger-Dougherty equation found for a wide variety of colloidal particles [50,51] :

$$\eta(c) = \eta_S \left(1 - \frac{c}{c_m}\right)^{-2} \qquad (3)$$

In Eq. 3, $\eta_S$ is the solvent viscosity and $c_m$ is related to the maximum-packing volume fraction through the relation $\phi_m = A c_m$, $A$ being a constant. According to the Krieger-Dougherty model [51], $c_m$ is the concentration at which the static viscosity diverges. Suspensions above $c_m$ exhibit the properties of yield stress materials, also called soft solids in the literature *i.e.* materials that can flow if they are mechanically subjected to shear stresses above a critical value. For Curosurf®, we found $c_m$ = 117 g L$^{-1}$. Assuming for the maximum-packing volume fraction the common value of 0.63 [52,53], one gets $A = 5.4 \times 10^{-3}$, in good agreement with the independently determined value of $A = 5.2 \times 10^{-3}$ from nanoparticle tracking analysis (**S6**). This second result confirms the value of the formulated Curosurf® volume fraction at $\phi$ = 0.42.

| Techniques | Concentration (g L$^{-1}$) | Static viscosity (mPa s) |
|---|---|---|
| Magnetic wire microrheology | 0 | 1.6 ± 0.4 |
| | 5 | 2.5 ±0.4 |
| | 20 | 3.4 ± 0.7 |
| | 40 | 6.0 ± 1.2 |
| | 50 | 8.5 ± 2.3 |
| | 70 | 16.5 ± 4.4 |
| | 80* | 24.5 ±5 / 17.9 ± 4.0 |
| Cone-and-plate rheology | 40 | 4.0 ± 0.5 |
| | 80 | 20.0 ±1.0 |

***Table I:*** *Values of Curosurf® static viscosity as a function of the concentration obtained from Magnetic Rotational Spectroscopy (MRS) and from cone-and-plate rheology. (*) The viscosity was determined from the average rotation frequency and from the oscillation amplitude.*

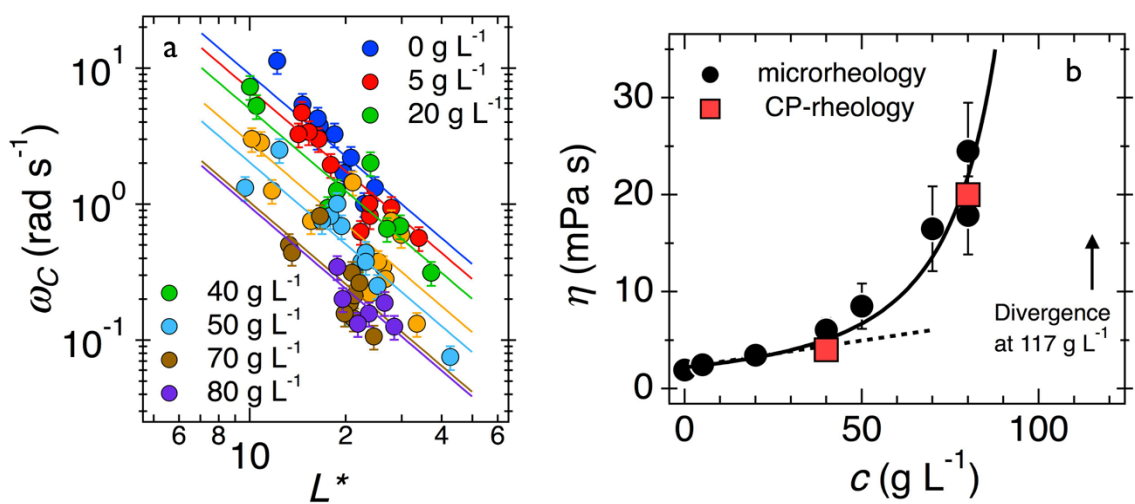

***Figure 5: a)*** *Critical frequency $\omega_C$ as a function of the reduced wire length $L^* = L/D\sqrt{g(L/D)}$ obtained for Curosurf® suspensions at different concentrations (T = 25 °C). Straight lines are least-square fits using Eq. 1.* ***b)*** *Concentration dependence of the Curosurf® static viscosity. The dashed line is obtained from the Einstein model while the continuous line was derived from the Krieger-Dougherty equation (Eq. 3) [50,51]. From the fitting, the vesicle maximum-packing concentration is determined ($c_m$ = 117 g L$^{-1}$).*

The results from Fig. 5b have several implications. The first one is that at the concentration at which Curosurf® is formulated (80 g L$^{-1}$), the viscosity is very sensitive to the lipid concentration.

Small errors in the determination of $\phi$ or $c$ can lead to large variation in the fluid viscosity. At 80 g L$^{-1}$, a 5% variation in the concentration results in a 50% increase in the viscosity, typically from 20 to 30 mPa s. The second outcome refers to the hypothesis made in the introduction about concentrating exogenous surfactant for treating RDS for adults. Under the current formulation conditions, the range over which the suspensions can be concentrated appears to be narrow, namely between 80 and 117 g L$^{-1}$. In this range the volume fraction approaches that of random close packing, leading to more viscous suspensions and strongly modified flow properties.

*III.2.4 – Onset of viscoelasticity at high concentration*
Figs. 6a and 6b display the wire oscillation amplitudes observed in Regime II as a function of the reduced frequency $\omega/\omega_C$ for Curosurf® 40 and 80 g L$^{-1}$. Wires between 20 and 80 µm were used in these measurements. At 40 g L$^{-1}$, $\theta_B(\omega)$ decreases with increasing frequency in accordance with the Newton constitutive equation prediction [49,54]. Away from the transition ($\omega \gg \omega_C$), the decrease goes roughly as $\theta_B(\omega) \sim \omega^{-1}$. In the $\theta_B(\omega/\omega_C)$-representation, the continuous line in the figure is obtained with no adjustable parameter and it accounts well for the data. At 80 g L$^{-1}$ however, $\theta_B(\omega)$ is found to deviate from the Newton prediction, an outcome that is interpreted as the onset of viscoelasticity (Fig. 6b). The continuous lines in the figure (from orange to green) depict Maxwell model predictions calculated for different relaxation times $\tau$.

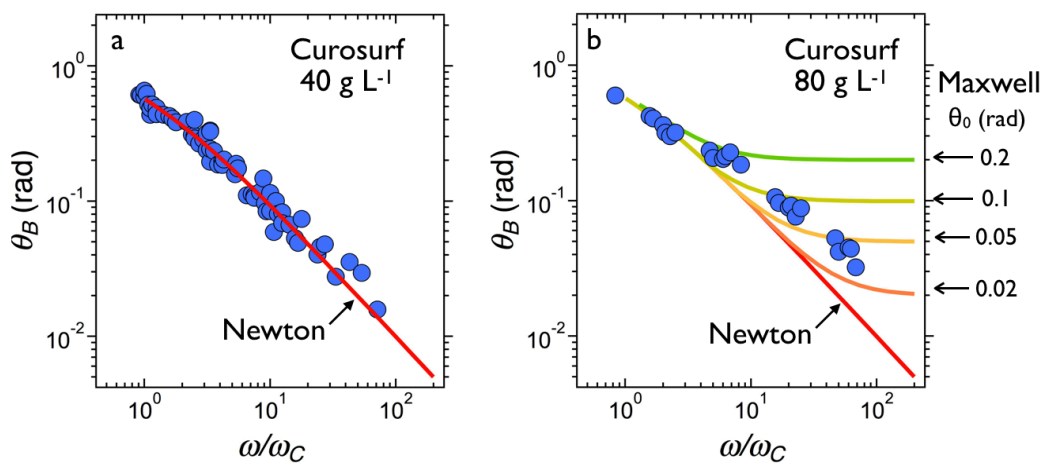

***Figure 6**: **a)** Oscillation amplitude $\theta_B(\omega/\omega_C)$ observed in Regime II for Curosurf® at 40 g L$^{-1}$ (T = 25 °C). The continuous line in red is the constitutive equation solution for Newtonian fluid. **b)** Same as **a)** for Curosurf® at 80 g L$^{-1}$. The continuous lines indicated by arrows are for the Maxwell model calculated for increasing relaxation times $\tau$ ranging from $0.01/\omega_C$ (orange) to $0.1/\omega_C$ (green).*

In macroscopic rheology, the Maxwell relaxation time $\tau$ is the ratio between the static viscosity and the elastic shear modulus ($\tau = \eta/G$). In the wire microrheology formalism, $\tau$ is related to the angle $\theta_0$ through the expression $\theta_0 = 2\omega_C\tau$ [49]. $\theta_0$ is defined as the high frequency limit of $\theta_B(\omega)$. In Fig. 6b, the oscillation amplitudes are outlined for $\theta_0$ between 0.02 to 0.2 rad. It can be seen in the figure that a single relaxation time Maxwell model is unable to account for the experimental data, indicating the existence of a broad relaxation time distribution for this fluid. According to the Maxwell model [41], the crossover from the viscous to elastic behavior occurs at $\omega\tau = 1/2$. The previous considerations allow to provide an estimate for the longest relaxation time in Curosurf® 80 g L$^{-1}$. An analysis on six different wires yields $\tau = 0.1$ s and from the relationship $G = \eta/\tau$ an elastic modulus $G = 0.2$ Pa. Such a low value also explains why the cone-and-plate rheometer was not able to measure $G'(\omega)$ properly. The above results demonstrate that the wire microrheology is a sensitive technique for measuring both viscous and elastic responses of soft materials.

# IV – Conclusion

Adapting neonatal surfactant replacement therapies to adults requires to reformulate the existing exogenous surfactants and increase their lipid concentration whilst maintaining unchanged their surface and flow properties. For neonates, lifesaving doses are in the range of 100 to 200 mg / kg of body weight, depending on the drugs administered. For delivered doses proportional to the weight of the patient for this type of treatment, volumes of liquid instilled in the lungs can reach 200 ml for an adult, a value that is considered high for such a medical intervention. When pulmonary surfactant is instilled into the trachea and into the airways, both interfacial and viscous properties are critical as they impact the propagating velocity of the liquid plug and the amount of deposited material on the nearby tissues [21,22]. There has been numerous studies devoted to pulmonary surfactant interfacial properties [14,15,35,55] and much less related to their rheological properties [27-29]. In particular accurate measurements of static shear viscosity (*i.e.* the viscosity extrapolated at zero frequency or zero shear rate) are missing for the most-used surfactant drugs available. In this work, we apply a microrheology technique where micron sized magnetic wires are remotely actuated thanks to the application of a rotating external field. As for the exogenous surfactant, we focus on Curosurf®, which is one of the most-used substitutes in European maternity hospitals for the treatment of respiratory distress syndrome. At the physiological concentration of 40 g L$^{-1}$, the static viscosity determined by magnetic microrheology amounts to $6.0 \pm 1.2$ mPa s, whereas at the clinically used concentration (80 g L$^{-1}$) it is $21.2 \pm 4.0$ mPas. Although Curosurf® is an expensive drug, we also studied these two samples using classical cone-and-plate rheometry and found static shear viscosities at $4.0 \pm 0.5$ and $20.0 \pm 1.0$ mPa s respectively, in good agreement with those of microrheology. Concerning the concentration behavior, we find that the Curosurf® viscosity varies according to the Krieger-Dougherty law over the whole concentration range. The Krieger-Dougherty behavior was observed on a wide variety of colloids [50] and it is characterized by an exponential increase of the viscosity and a divergence close to the liquid-to-gel transition. For Curosurf® this transition concentration is found at 117 g L$^{-1}$, which represent only a 50% increase compared to the concentration at which it is formulated. For the most concentrated specimen studied, we show the existence of viscoelasticity, with an elastic modulus of 0.2 Pa. These outcomes suggest that under the current formulation conditions, this surfactant substitute cannot be further concentrated and used for clinical treatments, as its viscosity becomes significantly large and the fluid becomes viscoelastic. Other strategies leading to a strong vesicle volume fraction reduction are required, such as the reduction of the vesicular size or a better control of the intra-vesicle interactions. The results found for Curosurf® could help in producing suspensions for respiratory distress syndrome adapted to adults. The present work demonstrates the potential of the wire-based active microrheology technique as an accurate tool to explore biological soft matter dynamics.

# Acknowledgments

We thank Armelle Baeza-Squiban, Victor Baldim, Yong Chen, Marcel Filoche, Daniel Isabey, Mélody Merle, Mostafa Mokhtari, Jesus Perez-Gil, Chloé Puisney, Milad Radiom, Nicolas Tsapis for fruitful discussions. Imane Boucema is acknowledged for letting us use the Anton Paar rheometer for the cone-and-plate rheology. ANR (Agence Nationale de la Recherche) and CGI (Commissariat à l'Investissement d'Avenir) are gratefully acknowledged for their financial support of this work through Labex SEAM (Science and Engineering for Advanced Materials and devices) ANR 11 LABX 086, ANR 11 IDEX 05 02. We acknowledge the ImagoSeine facility (Jacques Monod Institute, Paris, France), and the France BioImaging infrastructure supported by the French National Research Agency (ANR-10-INSB-04, « Investments for the future »). This research was supported in part by the Agence Nationale de la Recherche under the contract ANR-13-BS08-0015 (PANORAMA), ANR-12-CHEX-0011 (PULMONANO), ANR-15-CE18-0024-01 (ICONS), ANR-17-CE09-0017 (AlveolusMimics) and by Solvay.

# TOC Image

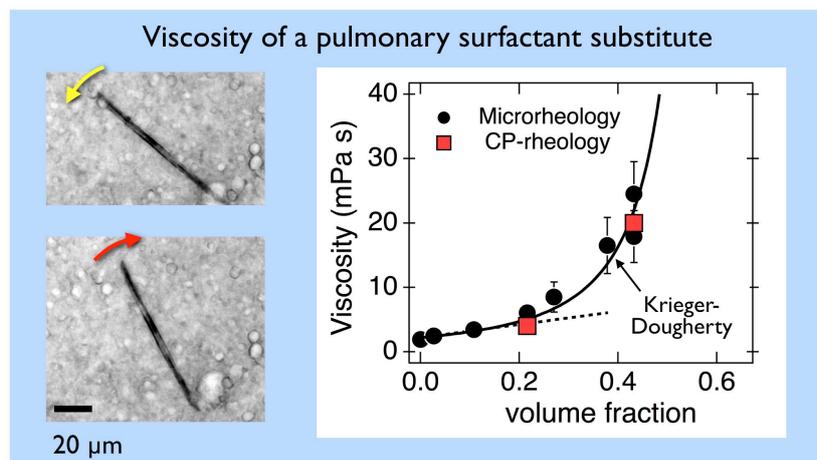